\documentstyle[epsfig]{mn}
\input{psfig}
\begin{document}
\def\ltsima{$\; \buildrel < \over \sim \;$}
\def\simlt{\lower.5ex\hbox{\ltsima}}
\def\gtsima{$\; \buildrel > \over \sim \;$}
\def\simgt{\lower.5ex\hbox{\gtsima}}

\title[The flat X-ray spectrum of the LINER NGC~1052]
{The flat X-ray spectrum of the LINER NGC~1052}

\author[M. Guainazzi, et al.]
{M. Guainazzi$^1$, L.A. Antonelli$^2$ \\ ~ \\
$^1$Astrophysics Division, Space Science Department of
ESA, ESTEC, Postbus 299, NL-2200 AG Noordwijk, The Netherlands \\
$^2$Osservatorio Astronomico di Roma, Via dell'Osservatorio 5, I-00040 Monteporzio Catone, Italy
}

\maketitle
\begin{abstract}
We report on ROSAT and ASCA observations of the LINER NGC~1052, which is
the first one where broad optical lines in polarized light have been
observed. The 2--10~keV spectrum is very flat, with a ${\rm \Gamma_{observed}
\simeq 0.1}$. A model where a nuclear source is - partly or totally -
obscured by a screen of matter with column density
${\rm \sim 10^{23}}$~cm$^{-2}$ is the most convincing explanation
for the observed flatness. This agrees with the hypothesis that
the LINERs are a population of low-luminosity AGN, to which the
Seyfert unification scenario applies.
The {\it intrinsic}
spectral index is still rather flat (${\rm \Gamma_{instrinsic}
\simeq 1.0}$--1.4), as observed in a few type-2 Seyferts
so far or predicted if the accretion occurs in
an advection-dominated flow.
\end{abstract}

\section{Introduction}

It is still a matter of debate whether Low Ionization Nuclear
Emission Line Regions
Galaxies (LINERs) are powered by intense starbursts, or by an
active galactic nucleus (AGN), or both. In some well-studied LINERs
the non-stellar nature is almost certain
({\it e.g.}: M81, M87, M104). However, for the majority of objects
the existence or relative amount of the AGN contribution remains unknown. 
Several facts point in favor of the idea that most LINERs
are low-luminosity AGN: the similarity in the host galaxy properties
and magnitude distribution (Ho et al. 1997), the discovery
of supermassive black holes in the central regions (Ho 1998),
and of broad wings of permitted optical lines (Ho 1999).

Studies in the
X-rays allow in principle
to probe the innermost regions of the galaxies up to a few
Schwarzschild radii around the putative central black hole
and to investigate the high-energy processes in the surrounding
environment. Actually, soft compact X-ray emission
has been detected in a handful of LINERs by the ROSAT HRI (Worral \&
Birinkshaw 1994; Koratkar et al. 1995; Fabbiano \& Juda 1997),
but the instrumental spatial resolution allowed
to limit the extension of the emitting region no
better than about a few hundred pc.
The ASCA observation of LINERs revealed that
a ``canonical'' simple model
is a general good description of the observed spectra, which is composed by
a hard power--law with photon index $\Gamma \simeq 1.7$--1.8 and a soft
excess of thermal origin, with typical temperatures $\simeq$1~keV
(see Ptak et al. 1998 and references therein).
Although the latter component closely resembles the non-thermal
power-law commonly observed in Seyfert 1s, the lack of iron K$_{\alpha}$
line emission (Terashima et al. 1998) argues against this identification.

If the LINERs are indeed a low-luminosity and nearby population of
Seyfert-like objects, one might expect that the Seyfert unification
scenario (see Antonucci 1993 for a review) applies to them as well
and that a population of LINERs exists, where the Broad Line
Region is hidden by optically thick matter. Actually, only a
fraction of the observed LINERs shows broad optical lines (Ho 1999).
A substantial support to this hypothesis has come from the recent discovery
of a broad H$_{\alpha}$ line in the polarized light spectrum of the
prototypical LINER NGC~1052 (Barth 1998), confirming previous weaker
claims (Ho et al. 1997). 

If the above
interpretation is correct, in some LINERs one might observe
X-ray spectra that are analogous to the ones typically observed in
Seyfert~2 galaxies: either 
a power-law spectrum is seen in transmission
through a ${\rm N_H \sim 10^{21-24}}$~cm$^{-2}$;
or, if the absorbing matter is optically thick to
Compton scattering (${\rm N_H \simgt 10^{24}}$~cm$^{-2}$),
only the reflection of the nuclear radiation
is detected. The last case has been observed in a handful
of so-called ``Compton-thick'' sources (Iwasawa et al. 1997;
Matt et al. 1996a, 1996b, 1997; Maiolino et al. 1998), including
the LINER NGC~6240 (Iwasawa \& Comastri 1998). 

NGC~1052 is an elliptical galaxy (3'$\times$2'.1) at a distance of about 
28~Mpc (or ${\rm z=0.0049}$, De Vaucouleurs 1991),
with ${\rm m_B = 11.8}$.
It is a LINER (Heckman 1980; Sadler 1987), located
in the group HG~44 (Huchra \& Geller 1982; Garcia 1993),
whose barycenter lies about 8' North from NGC~1052.
Optical and UV observations show an extended (20'') emission line 
region with shock heating emission spectrum 
(Fosbury et al. 1978; 1981) and an IR excess,
which is confined to a central region smaller than 2'' 
($\simeq$ 300 pc) and coincident with the optical nucleus. This suggests that 
IR emission could arise from dust heated by the non-thermal source observed in 
optical and UV (Becklin et al., 1982).
NGC~1052 is a
strong radio source and its radio spectrum
peaks at the GHz frequencies. The
spectrum is flatter than $\alpha$=-0.9 above 1 GHz and
is variable on a timescales of months (Disney \& Wall 1977).
VLBI observations showed a compact radio core with a 
diameter of 0.001 arcsec ($\simeq 0.14$~pc) and a radio halo
of about 21 arcsec (Fosbury et al. 1978).
Recently very luminous water masers have been detected toward this source 
(Claussen et al. 1998), which
are not aligned perpendicular to the radio jet, like in NGC~4258 
and NGC~1068, but rather lie along the jet at a distance of 0.07 pc from 
the central engine. This implies that masers emission in NGC~1052 should be 
associated with radio jet or radio continuum rather than with a molecular
region orbiting around the central engine (like in NGC~4258).

In this {\it Letter} we report still unpublished
results on observations of NGC~1052 taken by the ROSAT and ASCA
X-ray observatories. Statistical uncertainties are
quoted at 90\% level of confidence for one interesting parameter
(${\rm \Delta \chi^2 = 2.71}$); energies
are quoted in the source rest frame;
${\rm H_0 = 50}$~km~s$^{-1}$~Mpc$^{-1}$ and ${\rm q_0 = 0.5}$
are assumed, unless otherwise specified.

\section{Data preparation}

Data of NGC~1052 have been retrieved from the ASCA and ROSAT public
archives as screened event file lists, which have been pre-processed
and reduced according to standard criteria.
The log of the observations presented in this {\it Letter} is
reported in Table~\ref{tab2}
\begin{table}
\begin{footnotesize}
\centering
\caption{Observation log}
\label{tab2}
\vspace{0.05in}
\begin{tabular}{lcc} \hline \hline
Instrument & Date & Exposure time (ks) \\
HRI &04/08/94 & $\simeq$3.4 \\
HRI &02/10/95 & $\simeq$22 \\
PSPC & 23/07/1993 & $\simeq$14 \\
ASCA & 12/08/1996 & $\simeq$36 (SIS)\\
& & $\simeq$38 (GIS) \\ \hline \hline
\end{tabular}
\end{footnotesize}
\end{table}

The ROSAT/PSPC is a Position Sensitive Proportional Counter 
sensitive in the 0.1--2.0~keV band with a spatial resolution
of about 20'' for on-axis sources. 
Scientific products have been extracted from a circular
area of 1', which includes 99.5\% of the 0.2~keV source
photons (and an even higher fraction
at higher energies). The background spectrum has been
extracted from an annulus with internal radius of 3'
and external radius of 5' draw around the source.
In the analysis of PSPC spectra 
we eliminated PHA channels 
below 9 and above 200 due to calibration
uncertainties.  
ROSAT/HRI
is a High Resolution Imager with a spatial 
resolution of about 10'' and a limited (2-band) spectral 
response.
ASCA payload (Tanaka et al. 1994; Makishima et al. 1996)
is constituted by a pair
of Charge Couple Devices (Solid Imaging Spectrometers,
SIS0 and SIS1), sensitive
in the 0.6--9~keV band, and a pair of gas scintillating proportional counters
(Gas Imaging Spectrometers, GIS2 and GIS3), sensitive in the 0.7--10~keV band.
Only grade 0,1,2 and 4 of the SIS data have been used.
Scientific products have been extracted from circular
areas of 4', 3.25' and 6' around the apparent centroid of the
source, for the SIS0, SIS1 and GIS, respectively, including more than
99\% of the source photons. To remove
any possible contamination from HG~44, background
spectra have been extracted using the complement area of the same
chip where the source lies and from an annulus surrounding the
source extraction region and having its same area for the SIS and
GIS, respectively. The following
results are not substantially affected if different
regions of the detectors' field of views or spectra extracted from
blank sky fields are employed.
NGC~1052 is well above the background up to 8 and 10~keV in the
SIS and the GIS, respectively.
Spectra have been rebinned
in order to have at least 20 counts per channel, in order to ensure
the applicability of the $\chi^2$ test.
Total count rates are: $0.0519 \pm 0.0014$,
$0.0376 \pm 0.0012$, $0.0403 \pm 0.0013$,
$0.0436 \pm 0.0015$, and $0.0311 \pm 0.0017$~s$^{-1}$
for the SIS0, SIS1, GIS2, GIS3 and PSPC, respectively.

\section{Spatial and timing analysis}

\begin{figure}
\begin{center}
\epsfig{figure=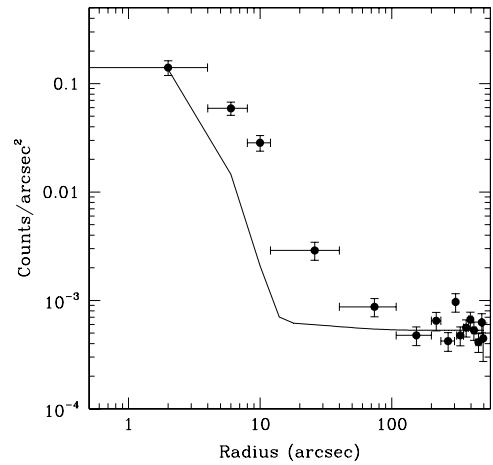,height=7.0cm,width=7.0cm}
\end{center}
\caption{Radial profile of the combined NGC~1052
HRI observations ({\it circles}), superimposed to the instrumental
PSF, as determined from the white dwarf HZ~43 profile
({\it solid line}). The
instrumental background (field of view average count rate
${\rm 5.4 \times 10^{-3}}$~arcsec$^{-2}$)
is accounted by a constant plateau in the PSF.
Each data point has a signal-to-noise ratio $5$}
\label{fig3}
\end{figure}

In Fig.~\ref{fig3} the NGC~1052 radial profile
extracted from the combined image of the two HRI observations
is compared with the instrumental
Point Spread Function (PSF). NGC~1052 shows extension up to 100'',
which is likely to be associated with the diffuse emission of the
galaxy. The count rate associated with the extended component
is 0.168~s$^{-1}$. It correspond to a
0.1--2~keV The luminosity of $\sim 3 \times 10^{40}$~erg~s$^{-1}$,
if one assumes a thermal bremsstrahlung with temperature
0.5~keV and seen through an absorbing column density of
$3 \times 10^{20}$~cm$^{-2}$.
No appreciable variability is detected,
either in the ASCA 0.5--1.5 and 1.5--10~keV or in the PSPC
0.5--2.4~keV light curves.
We will focus thereafter on the average spectral properties only.

\section{Spectral analysis}

Fitting separately the same model on the
spectra of the ASCA detectors yields consistent results within
the statistical uncertainties. Particular care has been adopted in
checking the consistency between the PSPC and the
ASCA spectra. Several authors have observed a systematic difference
in the slopes measured by ROSAT and ASCA in the overlapping energy band,
which ranges between 0.2 and 0.4 (cf. Fiore et al. 1994;
Iwasawa et al. 1998). Moreover, the source could have
varied between the two observations, both in flux and
in spectral shape, and the contribution of any
diffuse emission from the underlying galaxy
could be differently spread underneath the source
spot by the instrumental PSF. If we perform a simple power-law fit
with photoelectric absorption in the
0.8-2~keV band, the spectral indices turn out to be consistent
within the statistical uncertainties (${\rm \Gamma_{ROSAT} =
1.5\pm^{0.7}_{0.5}}$, ${\rm \Gamma_{ASCA} = 1.1 \pm 0.3}$).
Consistent results with those presented in this {\it Letter}
are obtained above 0.5~keV if the ASCA data alone are fitted.
At the light of the above considerations,
we have fitted the spectra
of all detectors simultaneously, allowing relative normalization
factors between different detector types as
free parameters in all the following fits. Their typical
values are always very close to 1 (${\rm N_{GIS}/N_{SIS} = 1.02}$--1.07, ${\rm
N_{PSPC}/N_{SIS} = 0.86}$--0.92).

\begin{figure}
\begin{center}
\epsfig{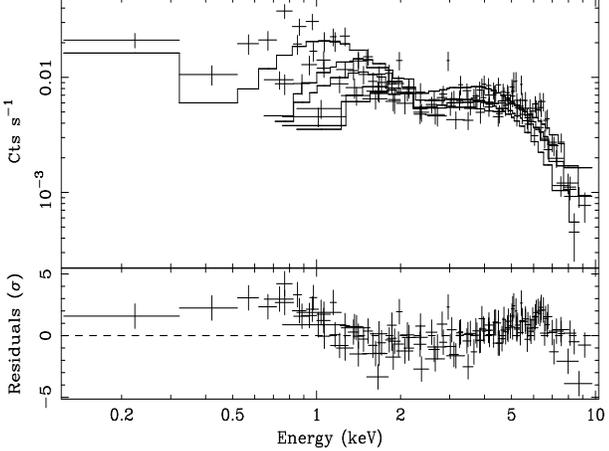}
\end{center}
\caption{Spectra ({\it upper panel}) and residuals in units of
standard deviations ({\it lower panel}) when a simple power-law
model with photoelectric absorption is applied. Each data point
has a signal-to-noise ratio $>5$. Energies in abscissa are in
the observer's frame}
\label{fig2}
\end{figure}

In Figure~\ref{fig2} the spectra and residuals are shown
when a simple power-law
model with photoelectric absorption by cold matter is applied. The
fit is rather bad (${\rm \chi^2_r = 1.59}$), mainly due to a prominent
soft excess below $\simeq$1.5~keV and an apparent
steepening of the spectrum above $\simeq$7~keV. A much better $\chi^2$
is obtained if a two-component model is adopted, where the soft excess
is parameterized either via a thermal optically thick (blackbody) or
thin ({\verb!mekal!}) plasma or through a broken power-law. In all cases,
an emission line is significantly required by the data
(${\rm \Delta \chi^2 = 37}$),
with centroid energy ${\rm E_c \simeq 6.35}$~keV, consistent with
K$_{\alpha}$ fluorescence from neutral or mildly ionized iron.
The 90\% upper limit on the Equivalent Width (EW) of a fluorescent
He-like (H-like) iron line is 170~eV (140~eV).
Table~\ref{tab1}
\begin{table*}
\begin{footnotesize}
\caption{
Best fit parameters and $\chi^2$ for the combined PSPC/ASCA spectral fits.
Explanations of the model abbreviations:
{\it wa}~=~photoelectric absorption;
{\it pc}~=~ partial covering; {\it po}~=~power-law;
{\it bp}~=~broken power-law
{\it bb}~=~blackbody; {\it mk}~=~MEKAL code (optically thin plasma);
{\it ga}~=~Gaussian emission line; 
{\it px}~=~Compton-reflection}
\label{tab1}
\vspace{0.05in}
\begin{tabular}{lcccccccccc} \hline \hline
\multicolumn {11}{c}{Power-law + soft excess} \\
Model & ${\rm N_H}$ & $\Gamma_{hard}$ & ${\rm kT}$ or ${\rm E_{break}}$ & $\Gamma_{soft}$ or ${\rm Z}$& ${\rm E_c}$ & EW & $\chi^2/$~dof & & & \\
& ($10^{20}$~cm$^{-2}$) & & (eV) & & (keV) & (eV) & & & &\\ 
{\verb!wa*(po+bb+ga)!} & $2.3 \pm^{1.6}_{0.8}$ & $0.11\pm^{0.07}_{0.08}$ & $165\pm^{18}_{15}$ & & $6.35 \pm^{0.07}_{0.08}$ & $220 \pm^{110}_{70}$ & 313.3/310 & & \\
{\verb!wa*(po+mk+ga)!} & $3 \pm 2$ & $0.11 \pm^{0.08}_{0.07}$ & $490 \pm 150$ & $0.07 \pm^{0.59}_{0.05}$ & $6.35 \pm^{0.06}_{0.07}$ & $240 \pm 80$ & 310.1/309 & & & \\
{\verb!wa*(bp+ga)!} & $4.1\pm^{1.6}_{2.0}$ & $0.15 \pm 0.07$ & $1.42 \pm^{0.16}_{0.12}$ & $1.9 \pm^{0.5}_{0.3}$ & $6.35 \pm^{0.07}_{0.06}$ & $240 \pm 80$ & 313.4/310 & & & \\ \hline
\multicolumn {11}{c}{Partial covering} \\
Model & ${\rm N_H^1}$ & $\Gamma$ & ${\rm N_H^2}$ & ${\rm C_f}$ & ${\rm E_c}$ & EW & $\chi^2/$~dof & & & \\
& ($10^{20}$~cm$^{-2}$) & & ($10^{22}$~cm$^{-2}$)& & (keV) & (eV) & & \\
{\verb!wa*pc(po+ga)!} & $2.6 \pm^{0.9}_{0.8}$ & $1.21 \pm 0.16$ & $10.3 \pm^{1.8}_{1.5}$ & $0.79 \pm 0.05$ & $6.36 \pm 0.06$ & $230 \pm^{80}_{90}$ & 296.9/310 & & & \\ \hline
\multicolumn {11}{c}{Pure reflection+soft excess} \\
Model & ${\rm N_H}$ & $\Gamma$ & ${\rm kT}$ & ${\rm Z}$ or ${\rm f_s \times (2 \pi/\Omega)}$& ${\rm E_c}$ & EW & $\chi^2/$~dof & & & \\
& ($10^{20}$~cm$^{-2}$) & & (eV) & & (keV) & (eV) & & & & \\
{\verb!wa*(px+po+ga)!} & $5.0 \pm^{0.9}_{0.8}$ & $2.16 \pm^{0.16}_{0.14}$ & & $0.0033 \pm_{0.0016}^{0.0011}$ & $6.4^{\dag}$ & $80 \pm^{70}_{30}$ & 290.8/312 & & & \\
{\verb!wa*(px+bb+ga)!} & $3.5 \pm^{1.3}_{0.9}$ & $2.43 \pm 0.09$ & $190 \pm 30$ & & $6.4^{\dag}$ &$90 \pm 70$ & 283.0/311 & & & \\
{\verb!wa*(px+mk+ga)!} & $4.3 \pm^{1.3}_{1.0}$ & $2.42 \pm^{0.10}_{0.09}$ & $620 \pm^{170}_{150}$ & $0.05 \pm^{0.20}_{0.04}$ & $6.4^{\dag}$ & $90 \pm^{100}_{70}$ & 278.0/309 & & & \\ \hline
\multicolumn {11}{c}{Transmission + soft excess} \\
Model & ${\rm N_H^1}$ & $\Gamma$ & ${\rm N_H^2}$ & ${\rm kT}$ & ${\rm f_s}$ or ${\rm Z}$ & ${\rm E_c}$ & EW & $\chi^2/$~dof & & \\
& ($10^{20}$~cm$^{-2}$) & & ($10^{22}$~cm$^{-2}$) & & (keV) & & (keV) & (eV) & & \\
{\verb!wa*[wa*(po+ga)+po]!} & $2.6 \pm^{0.9}_{0.7}$ & $1.22 \pm 0.16$ & $12 \pm 2$ & & $0.26 \pm^{0.15}_{0.09}$ & $6.36 \pm 0.06$ & $300 \pm 110$ & 297.0/310 & & \\
{\verb!wa*[wa*(po+ga)+bb]!} & $< 1.8$ & $0.49 \pm^{0.20}_{0.19}$ & $1.7 \pm^{1.1}_{0.8}$ & $260 \pm 50$ & & $6.36 \pm ^{0.06}_{0.07}$ & $260 \pm 90$ & 299.3/309 & & \\
{\verb!wa*[wa*(po+ga)+mk]!} & $2.5 \pm^{0.8}_{0.6}$ & $1.0 \pm^{0.5}_{0.4}$ & $9 \pm^6_3$ & $>9$& 1$^{\dag}$ & $6.35 \pm^{0.07}_{0.06}$ & $220 \pm 110$ & 296.0/309 & \\ \hline
\multicolumn {11}{c}{Mixed transmission + reflection} \\
Model & ${\rm N_H^1}$ & $\Gamma$ & ${\rm N_H^2}$ & ${\rm f_s}$ & ${\rm R}$ & ${\rm E_c}$ & EW & $\chi^2/$~dof & & \\
& ($10^{20}$~cm$^{-2}$) & & ($10^{22}$~cm$^{-2}$) & & & (keV) & (eV) & & & \\
{\verb!wa*[wa*(po+ga+px)+po]!} & $5.5 \pm^{1.7}_{1.6}$ & $2.1 \pm^{0.2}_{0.4}$ & $3.0 \pm^{1.9}_{1.8}$ & $0.5 \pm^{\infty}_{0.2}$ & $80 \pm^{\infty}_{60}$ & $6.38 \pm^{0.08}_{0.14}$ & $100 \pm^{90}_{80}$ & 280.6/309 & & \\ \hline \hline
\end{tabular}
\noindent
$^{\dag}$fixed
\end{footnotesize}
\end{table*}
reports the results and best-fit parameters for all the models
described in this section, which yield an acceptable $\chi^2$. All the
phenomenological descriptions are characterized by a very flat hard
spectral index
(${\Gamma \simeq 0.1}$), which is totally inconsistent with
that observed in
``canonical'' spectra of LINERs (Ptak et al. 1998).
The observed fluxes in the 0.5--2, 0.5--4.5 and 2--10~keV
bands are $\simeq 3.2 \times 10^{-13}$, 1.3 and
$3.5 \times 10^{-12}$~erg~cm$^{-2}$~s$^{-1}$,
respectively, corresponding to {\it unabsorbed} luminosities
of $3.6 \times 10^{40}$, 1.3 and $4.7 \times 10^{41}$~erg~s$^{-1}$,
respectively.
It is interesting
to note that the spectrum {\it cannot} be modeled with the composition of
thermal components alone, regardless of their nature and/or different types
of absorption. We have tried three different possible scenarios
(albeit not mutually exclusive on physical grounds) to account for
the observed flatness: (a) a ``partial covering'' of a nuclear
source; (b) a ``pure Compton-reflection'' model; (c) a ``transmission''
model, where the nuclear radiation is seen through a thick screen
of absorbing matter. In cases (b) and (c) a soft excess is still
required and has been parameterized with either a blackbody,
an optically thin plasma with free abundance or a scattering component,
{\it i.e.} a locally unabsorbed power-law with spectral index constrained
to be equal to that of the primary power-law.
A K$_{\alpha}$ fluorescent
iron emission line is required by all models as well.

The ``partial covering'' yields a very successful fit
(${\rm \chi^2_r = 0.95}$). A screen with
${\rm N_H \simeq 10^{23}}$~cm$^{-2}$, covering ${\rm \simeq 80\%}$ of the
primary source is required. The photon index of the primary
continuum is still very flat (${\rm \simeq 1.25})$.

The ``pure reflection'' scenario is also a very good parameterization of
the spectrum (${\rm \chi^2_r =}$0.90--0.93). The intrinsic spectral index
turns out to be even steeper than typically observed in Seyferts if the
soft excess is modeled through a thermal component
(${\rm \Gamma \simeq 2.3}$), but perfectly consistent with it
if the soft excess is due to scattering of the primary radiation.
However, the iron line is rather tiny in this scenario (EW $\simeq 80$--90~eV).

The ``transmission'' scenario is again a good model for the data
(${\rm \chi^2_r = 0.95}$--0.97). If the soft excess is modeled through
scattering or an optically thin plasma, the intrinsic spectral index is
again rather flat (${\rm \Gamma \simeq}$1.0--1.4) and seen through
a ${\rm N_H \sim 10^{23}}$~cm$^{-2}$ screen.
The unabsorbed luminosity of the nuclear component is $\simeq 5.6
\times 10^{41}$~erg~s$^{-1}$.
The iron line EW ranges between 100 and 400~eV.

We have also tried a ``mixed transmission+reflection'' scenario,
assuming that the nuclear primary
continuum is not a simple power-law, but is modified by Compton-reflection,
probably originating in
the putative relativistic Keplerian accretion disk spiraling
around the supermassive black hole. In this hypothesis the iron
line is expected to be significantly broadened (Matt et al. 1992; Tanaka
et al. 1995; Nandra et al. 1997), but the available statistics does not
allow us to set any constraint on the line profile.
An intrinsic Seyfert-like spectral index is obtained
(${\rm \Gamma \simeq 2}$), absorbed through a screen
of several ${\rm 10^{22}}$~cm$^{-2}$. However, this is got
at the cost of an unplausible high
relative normalization between the reflected and primary components
(${\rm R \simgt 20}$). The iron line is again rather weak
(${\rm EW \simeq 100}$~eV).
It is very difficult to imagine a geometry such
as to produce this huge amount of reflection when the primary radiation
is not completely suppressed in the ASCA bandpass by a Compton-thick
absorber, unless the nuclear emission is strongly beamed towards the
reflecting matter.

\section{Discussion and conclusions}

The X-ray spectrum of the LINER NGC~1052 is exceptionally flat. The
observed 2--10~keV photon index is ${\rm \Gamma \simeq 0.1}$.
The 2--10~keV luminosity in the same energy range
(${\rm \sim 5 \times 10^{41}}$~erg~s$^{-1}$)
is more than one order of magnitude higher than the expected
contribution
from unresolved Low Mass X-Ray Binaries within the
galaxy (Canizares et al 1987; Matsushita et al. 1994).
Models to
explain such a spectrum fall in two main classes: those which require
a highly absorbed (${\rm N_H \simeq 10^{23}}$~cm$^{-2}$) transmitted
power-law; (``partial covering''
and ``transmission'' scenarios);
and those where the 2--10~keV flux is dominated by Compton-reflection
(``pure reflection'' scenario).

In the former cases, the spectral index is still rather flat
(${\Gamma \simeq}$~1.0-1.4). Such values are distant from the average
observed in Seyferts (Nandra et al. 1997),
although not uncommon among type-2 objects
(Smith \& Done 1996; Turner et al. 1997).
Alternatively, such a flat index emerges naturally if accretion
occurs via an advection-dominated flow (ADAF, Narayan \& Yi 1995).
ADAF has been recently invoked to explain the emission from
low-luminosity AGN, thanks to the very low radiative conversion
efficiency that characterizes it.
It has been rather successful in modeling the Spectral Energy
Distributions of some low-luminosity AGN (see {\it e.g.}:
Narayan et al. 1998). It has been recently pointed out that
the canonical model significantly overestimates the radio
and sub-millimeter flux in a number of elliptical galaxies, hence suggesting
the existence of
mechanisms to suppress the emission from the central region at these
wavelengths (Di Matteo et al. 1998).
In the basic ADAF scenario the
X/$\gamma$-ray emission should be dominated by thermal bremsstrahlung
due to the very hot electron distribution in the disk, with typical
${\rm kT \sim 100}$~keV. In low-resolution, band-limited detectors such
a spectrum might be well approximated by a simple power--law
with ${\rm \Gamma \simeq 1.4}$.
Further observations covering the energy band up to a few hundreds keV
might be very valuable is assessing the viability of such a scenario,
and are nowadays possible thanks to the unprecedented sensitivity
of the Phoswitch Detector System on board BeppoSAX.

If the transmission scenario is
true in NGC~1052 (either in the Sakura-Sunayev or in the ADAF flavor),
two more ingredients are needed: {\it i)}
a screen of cold absorbing matter of ${\rm N_H \simeq
10^{23}}$~cm$^{-2}$, which could be provided by the
``torus'' envisaged by the unification scenario or (in the ADAF case)
by the cold outermost
regions of a geometrically bloated disk.
The observed iron line is consistent with
being produced by the same absorbing screen,
even if a contribution to the iron line
comes from a relativistic, X-ray illuminated disk
close to the nuclear black hole (note that an ADAF should not
intrinsically produce any fluorescent line);
{\it ii)} a soft component, either as partial covering, or as a thermal
emission peaking at several keV or as
an energy-independent $\simeq 25\%$ scattering of
the primary radiation.
The discovery of broad optical polarized lines advocates for scattering
indeed to occur in NGC~1052. However, there is evidence for an extended
soft X-ray component from the HRI data, which is likely to be
associated with the diffuse emission of the galaxy. The limited
spectral resolution of the HRI does not allow a self-consistent determination
of the luminosity of such a component. If one assumes that it is thermal in
origin with a temperature of 0.5~keV (comparable to
those typically observed in elliptical galaxies; cf. Fabbiano 1989),
the implied 0.1--2~keV luminosity is
${\rm \sim 3 \times 10^{40}}$~erg~s$^{-1}$.
From the observed correlation between the blue and the X-ray luminosities
in ellipticals (David et al. 1992),
one may expect a contribution to the 0.5--4.5~keV luminosity from the diffuse
galaxy emission of the same amount 
(David et al. 1992).
Of course different components
could be simultaneously present in the soft X-ray band, thus confusing
the interpretation of the data.
A detailed modeling of the soft excess in NGC~1052
requires experiments with much higher spatial resolution and/or
effective area.

Alternatively, a Compton reflection-dominated spectrum from a ``canonical''
Seyfert-like primary spectrum provides an equally good
parameterization of the data. However, in this case the
iron line is very faint, the EW being one order of magnitude
lower than
observed in Compton-thick Seyfert 2s so far (Iwasawa et al. 1997;
Matt et al. 1996b; Maiolino et al. 1998) and expected on theoretical
grounds (Ghisellini et al. 1994; Matt et al. 1996a). In principle, a strong iron
under abundance could resolve the discrepancy, although it is not
clear why this should be the case in the nucleus of an elliptical 
galaxy. Moreover, there is currently no observational
evidence for iron under-abundance in AGN
central regions. A more appealing solution may be that the
surface of the inner disk is ionized of such an amount
that the bulk of the
line flux is suppressed via resonant trapping (Matt et al. 1993).
We would therefore
be observing only a minority of the line flux, emerging from the
outermost and cold parts of the disk.
However, it is difficult to imagine a geometry where the inner
regions of the disk are visible, while the primary nuclear emission
is totally suppressed.

While a wide literature exists on the optical spectra of LINERs, the X-ray
spectrum of only a few objects has been studied in detail. In this
{\it Letter} we have presented evidence that an obscured (${\rm N_H
\sim 10^{23}}$~cm$^{-2}$) AGN, possibly accreting via an ADAF,
is the ultimate source powering NGC~1052. Undoubtedly, future
X-ray observations will contribute to further enlighten the connection
between LINER phenomenon and nuclear activity, and the nature of
the accretion processes occurring in LINERs.

\section*{Acknowledgments}

The authors acknowledge valuable discussions with F.Fiore and G.Matt,
which greatly improved the quality of the paper.
MG acknowledges an ESA research fellowship. This research has made use of
the NASA/IPAC Extragalactic Database, which is operated by the Jet
Propulsion Laboratory, Caltech, under contract with NASA, and of data
obtained through the HEASARC on-line service, provided by NASA/GSFC.

{}

\end{document}